\documentstyle[12pt,cite,epsf,here,amssym]{article}
\def \be {\begin{equation}}
\def \e {\end{equation}}
\def \bea {\begin{eqnarray}}
\def \ea {\end{eqnarray}}

\def \l {\lambda}
\def \no {\nonumber}

\def \sub {\scriptscriptstyle}
\newcommand{\To}[2]{\stackrel{#1}{\hbox to #2 pt{\rightarrowfill}}}

\def \today{\ifcase\month\or
  January\or February\or March\or April\or May\or June\or
  July\or August\or September\or October\or November\or December\fi
  \space\number\year}
\newcounter{Section}

\newcounter{Subsection}[Section]

\newcounter{Subsubsection}[Section]

\newcounter{Appendix}
\def \theAppendix{\Alph{Appendix}}
\newcommand{\app}[1]{\refstepcounter{Appendix}%
\centerline{\bf APPENDIX \theAppendix: \uppercase{#1}}\setcounter{equation}{0}%
\renewcommand{\theequation}{\Alph{Appendix}\arabic{equation}}}

\def\np#1#2#3{{\it Nucl.~Phys.\/}~{\bf B#1} (19#2) #3}

\def\pp{{\it preprint\/} }
\def\prd#1#2#3{{\it Phys.~Rev.\/}~{\bf D#1} (19#2) #3}

\def\prp#1#2#3{{\it Phys.~Rep.\/}~{\bf #1} (19#2) #3}

\def\zpc#1#2#3{{\it Z.~Phys.\/}~{\bf C#1} (19#2) #3}
\def\arn#1#2#3{{\it Ann.~Rev.~Nucl.~Part.~Sci.\/}~{\bf #1} (19#2) #3}
\begin{document}
\begin{flushright}
PITHA 97/02 \\ hep-ph/9701255 \\ January 1997
\end{flushright}
\begin{center}
\LARGE \bf{Energy Spectrum of Leptons from \boldmath $e^+e^- \to W^+W^-$ in the Presence of Strong \boldmath  $W^+_{\sub{L}}W^-_{\sub{L}}$ Interaction}
\end{center}
\begin{center}
\sc Anja Werthenbach \footnote{\footnotesize
Electronic address: anja@physik.rwth-aachen.de} and
L.\,M.~Sehgal\footnote{\footnotesize Electronic address: sehgal@physik.rwth-aachen.de}
\\ \it Institut f\"ur Theoretische Physik (E), RWTH Aachen\\
D-52056 Aachen, Germany
\end{center}
\vspace{1.0cm}
\thispagestyle{empty}
\centerline{\bf ABSTRACT}
\begin{quotation}
The energy spectrum of leptons, produced by the decay of one or both $W$'s in the reaction $e^+e^- \to W^+W^-$, is a significant probe of the helicity structure of this process. We calculate the energy spectrum $d \sigma /d E$ of a single decay lepton, as well as the two-dimensional energy distribution $ d \sigma/d E_+dE_-$. We then consider a possible strong final state interaction in the longitudinally polarized state of the $W^+W^-$ system, parametrized by a $\rho $-like resonance in the region of $1.8$ TeV. We show that such a resonance produces measurable effects in the lepton spectrum at $\sqrt{s}=500$ GeV. The results are compared with those obtained from a non-resonant interaction approximated by a phase factor $e^{i \delta}$ in the $l=1$ part of the $W^+_{\sub{L}}W^-_{\sub{L}}$ amplitude.
\end{quotation}
\setcounter{footnote}{0}
%
%
\newpage
The differential cross section of the reaction $e^+e^- \to W^+W^-$ in the standard model has an intricate helicity structure, that is reflected in the angular distribution of this process \cite{hagi}:
\be
\label{cross}
\frac{d \sigma}{d\! \cos\! \Theta} = \sum _{\l_-, \l_+ }  {\left( \frac{d \sigma}{d\! \cos\! \Theta} \right) }_{\l_- \l_+} \qquad .
\e
Here $ \Theta$ is the angle of the $W^-$ relative to the $e^-$ beam, and $\l_- (\l _+) $ are the $W^- (W^+)$ helicities. Since the amplitude of the reaction is determined by $\gamma$- and $Z$- exchange in the $s$-channel and $\nu$-exchange in the $t$-channel, the helicity structure of the process is sensitive to the gauge couplings $ \gamma WW$ and $ZWW$.

A measurement of the angular distribution (\ref{cross}) requires that at least one of the $W$'s decays into two jets, whose momenta can be measured in order to determine the $W^-W^+$ production axis. However, even in the absence of information about the production angle, very significant information about the helicity structure of the process can be obtained from the energy distribution of the leptons created by the decay of one (or both) $W$'s: $W^{\pm} \to \ell^{\pm} \nu$. It is important to note that the energy of the decay lepton in the laboratory (or $e^+e^-$ center-of-mass) frame is directly related to the polar angle of emission of the lepton with respect to the $W$-direction, in the $W$-rest frame:
\bea
\label{energy}
E_- &=& \frac{1}{2} E(1-\beta \cos \theta_-) \no \\
E_+ &=&  \frac{1}{2} E(1-\beta \cos \theta_+)
\ea
with $\beta=(1- 4M_{\sub{W}}^2/s)^{\frac{1}{2}}$, $E$ being the beam energy, $E_{\mp}$ being the energy of $\ell_{\mp}$ and $\theta _{\mp}$ being the angle between the direction of $W^{\mp}$ and $\ell_{\mp}$.
It is convenient to define dimensionless variables $X_{\mp}$ related to $E_{\mp}$ by
\bea
\label{X}
X_{\mp}&=& \frac{1}{E\beta} \left[ E_{\mp} -\frac{E}{2} (1-\beta)\right]  \no \\
0&< &X_{\mp}<1
\ea
Then the two-dimensional spectrum in the energy variables $X_{\mp}$ is
\be
\label{spectrumX}
\frac{d \sigma}{dX_-dX_+}= B_{\ell}^2 \sum_{\l_-,\l_+} \sigma_{\l_-\l_+} {\cal D}_{\l_-}(X_-){\cal D}_{\l_+}(1-X_+)
\e
where
\be
\label{sigma}
\sigma_{\l_-\l_+}=\int_{-1}^{+1} {\left( \frac{d\sigma}{d\! \cos\! \Theta} \right)}_{\l_-\l_+}\!\! d\! \cos\! \Theta
\e
and
\be
\label{decay}
{\cal D}_{\l}(X) = \left\{ \begin{array}{l@{\quad: \quad}l} 3X^2 &\l =+1 \\ 6X(1-X)& \l=\;\;\,0 \\ 3(1-X)^2& \l =-1 \end{array} \right.
\e
$B_{\ell}$ being the leptonic branching ratio.

The nine helicity cross sections $\sigma_{\l_-\l_+}$ are listed in the Appendix, where for generality, we give the results separately for $e^-_{\sub{L}}$ and $e^-_{\sub{R}}$ beams. The energy spectrum  $d \sigma / dX_+dX_- $ is thus a linear combination of the helicity cross sections $\sigma_{\l_-\l_+}$ (integrated over the $W^-W^+$ production angle), and affords a test of the helicity structure of this reaction, without the need to reconstruct the $W^-W^+$ production axis. If only one of the two $W$'s decays leptonically (say, $W^- \to \ell^- \bar{\nu}$), the energy spectrum is
\be
\label{oneenergy}
\frac{d \sigma}{d X_-} =B_{\ell} \sum_{\l_-=\pm,0}\sigma_{\l_-} {\cal D}_{\l_-}(X_-)
\e
the function $\sigma_{\l_-}$ being given by $\sigma_{\l_-} = \sum_{\l_+}\sigma_{\l_-\l_+}$. The energy spectrum for unpolarized $e^-e^+$ beams obtained by averaging the results for $e^-_{\sub{L}} $ and $e^-_{\sub{R}} $ given in the Appendix, agrees with that calculated by Dicus and Kallianpur \cite{dicus}.

We now contemplate a possible strong final state interaction in the $W^-W^+$ system, and its influence on the secondary lepton spectrum. Specifically, we consider the idea that longitudinally polarized $(\l=0)$ $W^-W^+$ pairs may have interactions analogous to those in the $\pi \pi$ system, producing, among other things, resonances in various partial waves \cite{chanowitz}. A concrete scenario for strong $W^-(\l =0)W^+(\l =0)$ interaction is provided by the technicolor model \cite{techni}, in one version of which a $\rho$-like resonance is predicted with a mass $m_{\rho} \sim 1.8 $ TeV and width $\Gamma_{\rho} \sim 0.5$ TeV. A final state interaction of this type has immediate implications for the helicity cross section $\sigma_{00}$, and affects the energy spectrum of the decay leptons. It is this effect that we wish to investigate.

The differential cross section $(d\sigma/d \cos \Theta)_{00}$ corresponding to the helicity state $W^-(\l=0)W^+(\l =0)$ can be written formally as
\be
\label{00}
{\left( \frac{d \sigma}{d\!\cos\! \Theta}\right)} _{00} \sim \big| {\cal M}_{\gamma + Z}^{00} + {\cal M}_{\nu }^{00} \big| ^2
\e
where ${\cal M}_{\gamma + Z}^{00}$ and ${\cal M}_{\nu }^{00}$ represent the matrix elements corresponding to $\gamma + Z$ exchange in the $s$-channel and $\nu $-exchange in the $t$-channel \cite{hagi}. In the presence of a final state interaction affecting the $l=1$ partial wave of the $W^-_{\sub{L}}W^+_{\sub{L}}$ system, the amplitude ${\cal M}_{\gamma + Z}^{00}$, which is pure $l=1$, is modified according to
\be
\label{gZmodi}
{\cal M}_{\gamma + Z}^{00} \to \Omega(s) {\cal M}_{\gamma + Z}^{00}
\e
where $\Omega(s)$ is a form factor representing the final state interaction. The neutrino-mediated amplitude ${\cal M}_{\nu }^{00}$ contains $l=1$ as  well as $l>1$ components, and is therefore modified as follows \cite{orsay}:
\be
\label{numodi}
{\cal M}_{\nu }^{00} \to \Omega (s) {\cal M}_{\nu }^{00}(l=1) + {\cal M}_{\nu }^{00} (l>1)
\e
where
\bea
\label{l=1,l>1}
{\cal M}_{\nu }^{00}(l=1) &=& \frac{3}{4} \int _{-1}^{+1} d\! \cos\! \Theta\, \sin\! \Theta {\cal M}_{\nu }^{00}(\cos\! \Theta) \no \\
{\cal M}_{\nu }^{00}(l>1) &=& {\cal M}_{\nu }^{00} - {\cal M}_{\nu }^{00}(l=1) \quad .
\ea
Thus equation (\ref{numodi}) may be written equivalently as
\be
\label{nuModi}
{\cal M}_{\nu }^{00} \to {\cal M}_{\nu }^{00} +\left( \Omega(s) -1 \right) {\cal M}_{\nu }^{00}(l=1)  .
\e
The function $\Omega(s)$ may be regarded as an Omnes function related to the $W^-_{\sub{L}}W^+_{\sub{L}}$ $p$-wave scattering phase shift by \cite{barklow}
\be
\Omega(s) = \exp \left[ \frac{s-4M_{\sub{W}}^2}{\pi}\: P\!\! \int_{4 M_{\sub{W}}^2}^{\infty} \frac{\delta (s')d s'}{(s'-4M_{\sub{W}}^2)\;(s'-s-i\epsilon)}
 \right]   .
\e
For the purpose of the present investigation, we approximate $\Omega(s)$ by a Breit-Wigner form factor
\bea
\label{BW}
\Omega(s) &=& \frac{(s-m_{\rho}^2)}{s-m_{\rho}^2+i\Gamma_{\rho} m_{\rho} \left(\frac{\beta}{\beta_{\sub{V}}} \right)^3} \no \\
\beta_{\sub{V}}&=&{ \left( 1-\frac{4 M_{\sub{W}}^2}{m_{\rho}^2}\right)}^{\frac{1}{2}}
\ea
normalized to $\Omega(s=4M_{\sub{W}}^2)=1$. The parameters $m_{\rho}$ and $\Gamma_{\rho}$ are chosen to be those of the techni-$\rho$ meson $(m_{\rho}=1.8 \mathrm{\,TeV},\Gamma_{\rho}=0.5 \mathrm{\,TeV})$ predicted in some versions of the technicolor model. (A more elaborate parametrization of the form factor $\Omega(s)$, based on an analogy with the Gounaris-Sakurai formula for $\pi \pi$ scattering, is also possible, and has been used in Ref. \cite{orsay}.) We also consider for comparison, a non-resonant final state interaction characterised by a simple phase factor $\Omega(s)= e^{i \delta(s)}$ where $\delta$ is taken to be $ 10^{\circ}$ at $\sqrt{s}=500$ GeV. The modification of the ${\cal M}^{00}$ matrix elements specified in Eq.(\ref{gZmodi}) and (\ref{numodi}) changes the helicity cross section $\sigma_{00}$ given in the Appendix, leaving the others unchanged.

\begin{figure}[h]
\vskip -3cm
\centerline{\epsfysize=24cm\epsffile{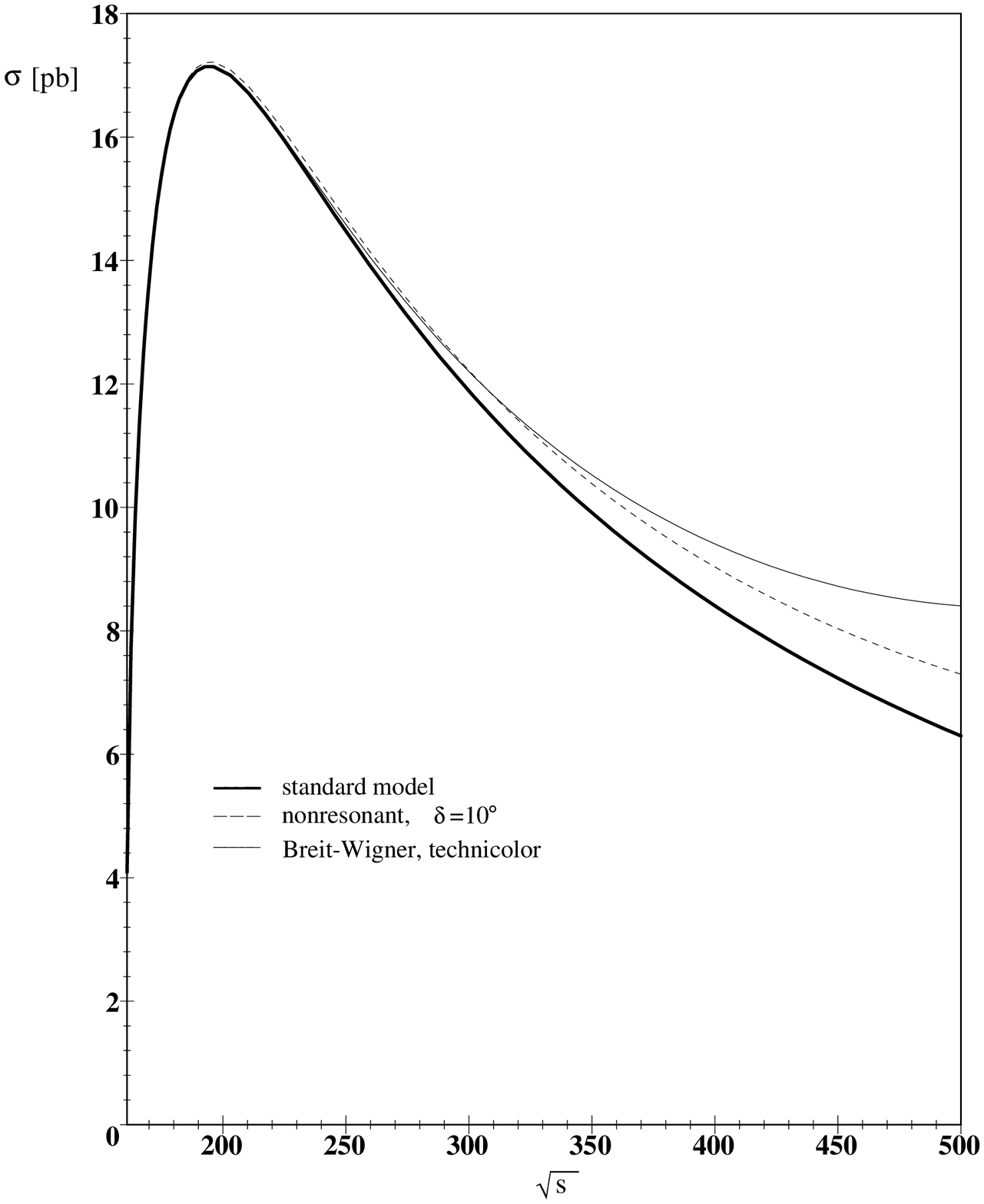}}
\caption{}\label{total}
\end{figure}

\begin{figure}[h]
\vskip -3cm
\centerline{\epsfysize=24cm\epsffile{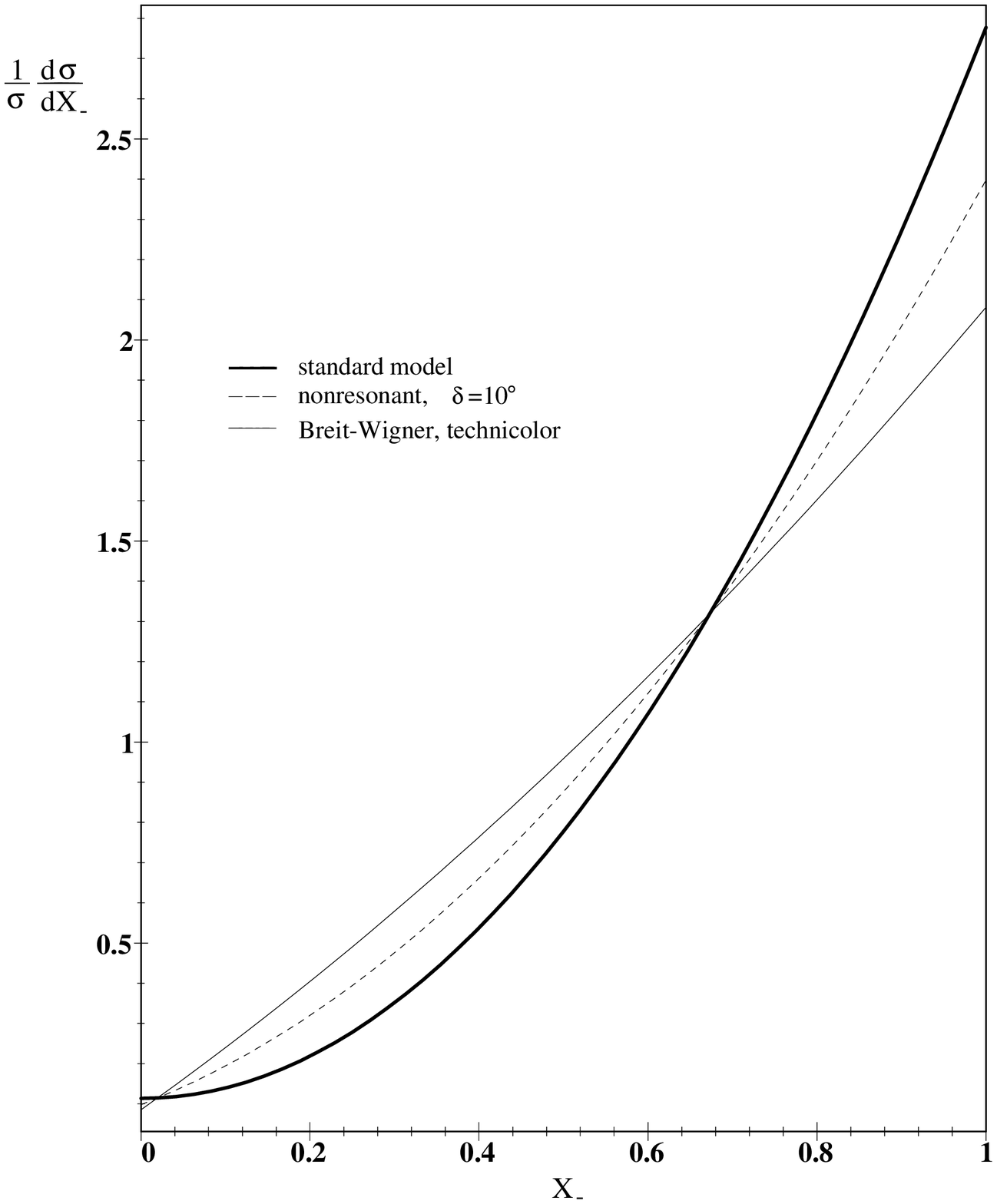}}
\caption{}\label{leptonenergy}
\end{figure}

The effect of final state interaction on the total cross section for $e^-e^+ \to W^-W^+$ is shown in Fig.\ref{total}.
At $\sqrt{s}=500$ GeV, the cross section is enhanced by about $35 \%$ in the case of a $\rho$-like resonance, and by about $15\%$ in the case of a non-resonant form factor with a phase shift of $10^{\circ}$. The effect on the lepton energy spectrum $d \sigma /d X$ is shown in Fig.\ref{leptonenergy} , for both types of form factor. Note, particulary, the enhancement in the region $0.1<X<0.6$ and the suppression for $X \gtrsim 0.7$.

\begin{figure}[h]
\vskip -3cm
\centerline{\epsfysize=24cm\epsffile{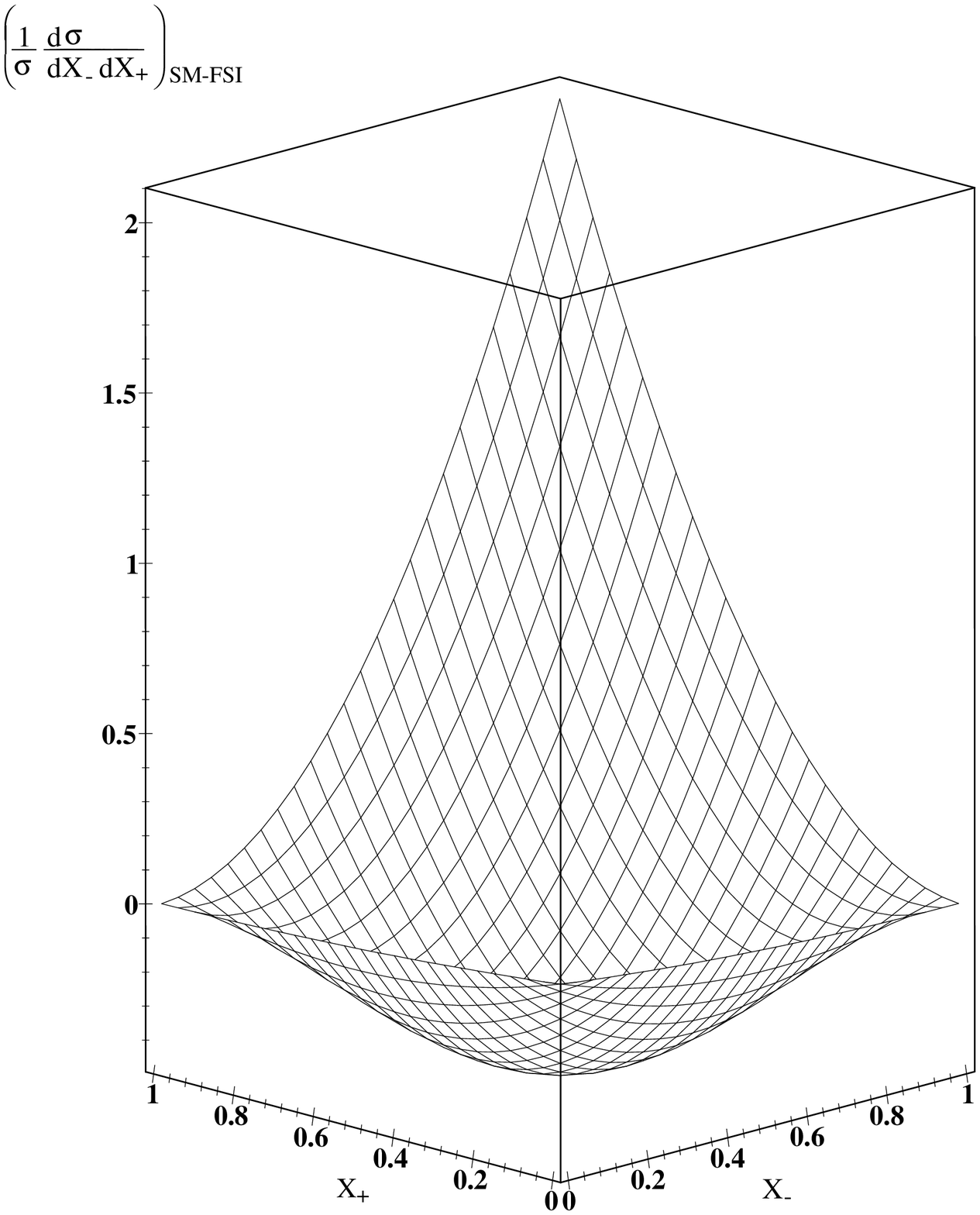}}
\caption{}\label{twodim}
\end{figure}

Finally, we show in Fig.\ref{twodim} the distortion in the normalized two-dimensional energy spectrum $\frac{1}{\sigma}d\sigma/ dX_-dX_+$ due to final state interaction. 
This distortion has its origin in the characteristic dependence $d \sigma_{00}/dX_+dX_- \sim X_+(1-X_+)X_-(1-X_-)$ given in Eq.(\ref{decay}). We remark once more, that the energy spectra shown in Fig.\ref{leptonenergy} and Fig.\ref{twodim} refer to cross sections integrated over all $W^-$ production angles, and therefore do not require measurements of jets to establish the $W$ momentum direction.

We conclude that the energy spectra of secondary leptons emerging from the reaction $e^-e^+ \to W^-W^+$ are sensitive to the helicity structure of the $W^-W^+$ final state. A final state interaction of the type expected in technicolor models, produces changes in these spectra that would be detectable in an $e^-e^+$ collider at $\sqrt{s}=500$ GeV, assuming an integrated luminosity of $10^4 \mathrm{\,pb}^{-1}$, corresponding to $\sim 2 \times 10^4 \,\,W^-W^+$ events with at least one charged decay lepton \cite{collider}. While specific versions of the technicolor model may be difficult to reconcile with the high precision tests at the $Z^0$ resonance \cite{tests}, the general idea of pion-like dynamics in the longitudinally polarized $WW$ system remains an interesting alternative to the standard model of electroweak symmetry-breaking \cite{bess}. Our analysis indicates that the energy spectrum of leptons from $e^-e^+ \to W^-W^+$ at c.m. energies $\sqrt{s} \simeq 500$ GeV could be an effective probe of this dynamics.

\vspace{1.5cm}

%
%
%
\app{Helicity Cross Sections}\label{app1}
\medskip
Helicity cross sections $\sigma_{\l_-\l_+}$ for $e^-_{\sub{L}}e^+_{\sub{R}} \to W^-_{\l_-}W^+_{\l_+}$ in the standard model, in units $m_{\sub{Z}}\equiv 1$, $m_{\sub{Z}}$ being the mass of the $Z^0$ boson:

\bea
\!\!\!\!\!\!\!\! \sigma_{00}&=&C_1 \frac{1}{(\beta^2 -1)^2 (s-1)^2}  \left[ \left( -3s\beta^{12} +3s^2\beta^{12}-6sx\beta^{12}+6x\beta^{12}-15s^2\beta^{10} +9s\beta^{10} \right. \right. \no \\
&&\mbox{}-42x\beta^{10} +6\beta^{10} +42sx\beta^{10}-30s\beta^8+108x\beta^8-12\beta^8+42s^2\beta^8 -108 sx\beta^8 \no \\
&&\mbox{}-78s^2\beta^6 -132x\beta^6+132 sx\beta^6-12\beta^6+90s\beta^6 +87s^2\beta^4-78sx\beta^4+48\beta^4 \no \\
&&\left. \mbox{}+78x\beta^4+135s\beta^4-42\beta^2-51s^2\beta^2 +18sx\beta^2+93s\beta^2-18x\beta^2+12-24s+12s^2\right) L \no \\
&&\mbox{}+\left( -8sx\beta^{11} +12s\beta^{11}+32x^2\beta^{11} -4s^2\beta^{11} -24x\beta^{11} -24s^2\beta^9 +96x\beta^9 -24\beta^9 \right. \no \\ 
&&\mbox{}-192x^2\beta^9+96sx\beta^9 -152s\beta^7+72\beta^7 +16x\beta^7+152 s^2\beta^7 +288x^2\beta^7-304sx\beta^7  \no \\
&&\mbox{}+384s\beta^5-288x\beta^5-264s^2\beta^5 -120 \beta^5 +288sx\beta^5+152\beta^3 -72sx\beta^3 +188s^2\beta^3  \no \\
&&\left. \left. \mbox{}+72x\beta^3-340s\beta^3-48\beta+96s\beta-48s^2\beta \right) \right] \no  \\
\!\!\!\!\!\!\!\!\sigma_{++}&=& C_1 \frac{1}{(s-1)^2} \frac{1}{2} \left[  \left( 6x\beta^8 +3s^2\beta^8 -3s\beta^8 -6sx\beta^8 +18sx\beta^6 -9s^2\beta^6 +9s\beta^6\right. \right. \no \\
&&\mbox{}-18x\beta^6 -21s\beta^4 +18x\beta^4 +6\beta^4+15s^2\beta^4 -18sx\beta^4 -6x\beta^2+6sx\beta^2 \no \\
&&\left. \mbox{} -15s^2\beta^2+27s\beta^2 -12\beta^2 +6 -12s+6s^2 \right) L \no \\ 
&&\left( -24 x\beta^7+4s^2\beta^7-40sx\beta^7+64x^2\beta^7+12s\beta^7+64sx\beta^5+32s\beta^5-32s^2\beta^5-64x\beta^5 \right. \no \\
&&\left.\left. \mbox{} +40\beta^3+24x\beta^3+52s^2\beta^3-24sx\beta^3-92s\beta^3+48s\beta-24\beta-24s^2\beta \right) \right] \no \\
\!\!\!\!\!\!\!\!\sigma_{--}&=&\sigma_{++} \no \\
\!\!\!\!\!\!\!\!\sigma_{-+}&=&C_1 \left[ \left( 3\beta^6+9\beta^5 +9\beta^4+6\beta^3+9\beta^2+9\beta +3 \right) L+\left( -12\beta^5 -36\beta^4-40\beta^3-36\beta^2-12\beta \right) \right]  \no \\ 
\!\!\!\!\!\!\!\!\sigma_{+-}&=&C_1 \left[ \left( 3\beta^6-9\beta^5 +9\beta^4-6\beta^3+9\beta^2-9\beta +3 \right)L+ \left( -12\beta^5 +36\beta^4-40\beta^3+36\beta^2-12\beta \right) \right] \no \\
\!\!\!\!\!\!\!\!\sigma_{0\pm}&=&C_1\frac{1}{ (s-1)^2} \Bigg[ \left(\pm 3\beta (s-1) (-s\beta^2 +2x\beta^2 +s-1) \left\{  (\beta^4+2\beta^2-3)L +(-4\beta^3 +12\beta) \right\} \right) \Bigg. \no \\
&&+\left[ \frac{1}{(\beta^2-1)}\left( 3s\beta^8 -18x\beta^8-6s^2\beta^8+3\beta^8+18sx\beta^8-3\beta^6+12s^2\beta^6 -9s\beta^6 -30sx\beta^6\right.  \right. \no \\
&&\mbox{}+30x\beta^6-3s\beta^4+3\beta^4 -6x\beta^4 +6sx\beta^4-12s^2\beta^2-9\beta^2-6x\beta^2+6sx\beta^2+21s\beta^2+6 \no \\
&&\left. \mbox{}+6s^2-12s \right) L+\left(-12\beta^7-128x^2\beta^7 +56sx\beta^7 +72x\beta^7 -12s\beta^7-8s^2\beta^7-8s^2\beta^5   \right.  \no \\
&&+32s\beta^5 +32x\beta^5 -24\beta^5-32sx\beta^5 +40s^2\beta^3-68s\beta^3 +24x\beta^3  -24sx\beta^3  +28\beta^3  \no \\
&&\Bigg. \Bigg. \left.  \mbox{}+48s\beta -24\beta-24s^2\beta \right) \Bigg] \Bigg]   \no \\
\!\!\!\!\!\!\!\!\sigma_{\mp0}&=&\sigma_{0\pm} \no \\ 
\ea
with $C_1 =e^4/{(24 x^2\beta^4 32\pi s)}$, $L=\ln\left( \frac{(1+\beta)^2}{(1-\beta)^2}\right)$ and $x=\sin^2 \theta_{\sub{W}}$. \\ 
\medskip

Helicity cross sections $\sigma_{\l_-\l_+}$ for $e^-_{\sub{R}}e^+_{\sub{L}} \to W^-_{\l_-}W^+_{\l_+}$ in the standard model, in units $m_{\sub{Z}}\equiv 1$:

\bea
\!\!\!\!\!\!\!\!\sigma_{00}&=& C_2 (\beta^2-3)^2 \frac{1}{(1-\beta^2)^2} \no \\
\!\!\!\!\!\!\!\!\sigma_{++}&=&\sigma_{--} = C_2  \no \\
\!\!\!\!\!\!\!\!\sigma_{0+}&=&\sigma_{0-}=\sigma_{+0} = \sigma_{-0} = 4  C_2 \frac{1}{(1-\beta^2)} \no \\
\ea
with $C_2 =\frac{4}{3} (e^4 \beta^3)/ (32\pi s(s-1)^2)$.

%
%

%
%
\newpage
\centerline{\bf FIGURE CAPTIONS}
\begin{enumerate}
\item[\bf Figure 1] Effect of final state interaction on the total cross section for $e^-e^+ \to W^-W^+$. 
\item[\bf Figure 2] Influence of final state interaction on the energy spectrum of the charged decay lepton in $e^-e^+ \to W^-W^+$, $W^- \to \ell^- \bar{\nu}$.
\item[\bf Figure 3] Shown is the correction to the normalized cross section $(1/ \sigma)( d \sigma / dX_-dX_+)$ (standard model minus modified) for $e^-e^+ \to W^-W^+ \to \ell^- \ell^+ \nu \bar{\nu}$  in the presence of final state interaction corresponding to a techni-rho resonance.
\end{enumerate}
\vspace{3cm}
%
%


\newpage
\begin{thebibliography}{99}
%
%
\bibitem{hagi} K. Hagiwara, R.\,D. Peccei, D. Zeppenfeld and K. Hikasa, \np{282}{87}{253} \\
K.\,J.\,F. Gaemers and G.\,J. Gounaris \zpc{1}{79}{259} \\
D. Chang, W.\,Y. Keung and I. Phillips \prd{48}{93}{4045}
\bibitem{dicus} D.\,A. Dicus and K. Kallianpur, \prd{32}{85}{35}
\bibitem{chanowitz} M.\,S. Chanowitz, \arn{38}{88}{323}
\bibitem{techni} E. Farhi and L. Susskind, \prp{74}{81}{277} \\
K. Lane and M.\,E. Peskin, "An Introduction to Weak Interaction Theories with Dynamical Symmetry Breaking," 15th Rencontre de Moriond (1980)
\bibitem{orsay} F. Iddir, A.\,Le Yaouanc, L. Oliver, O. P\`ene and J.-C. Raynal, \prd{41}{90}{22}
\bibitem{barklow} M. Peskin, in Physics in Collisions IV, Santa Cruz, CA, 1984, edited by A. Seiden (\'Editions Fronti\`eres, Gif-Sur-Yvette, France, 1984)\\
T.\,L. Barklow, in Proceedings of the Workshop -  Annecy, Gran Sasso, Hamburg, Feb. 95 to Sep. 95, DESY 96-123D, edited by P.\,M. Zerwas, p. 263
\bibitem{collider} Proceedings of the Workshop - Munich, Annecy, Hamburg, Nov. 92 to April 93, DESY 93-123C , edited by P.\,M. Zerwas
\bibitem{tests} G. Altarelli, Precision tests of the standard model at $Z^0$ \pp{hep-ph/9611239}
\bibitem{bess} N.\,Di Bartolomeo and R. Gatto, Breaking Electroweak Symmetry Strongly \pp{hep-ph/9404264} to appear in the Memorial Volume for Professor R. Marshak, edited by E.\,C.\,G.  Sudarshan, World Scientific Publishing Company  \smallskip \\
K. Hagiwara, T. Hatsukano, S. Ishihara and R. Szalapski, Probing nonstandard bosonic interactions via $W$-boson pair production at lepton colliders \pp{hep-ph/9612268} and references therein



\end{thebibliography}
\end{document}